\documentclass[11pt,a4paper]{article}

\usepackage{graphicx}  
\usepackage{amsfonts}

\title{Gilkey-de Witt heat kernel expansion and zero modes}

\author{A. Alonso Izquierdo$^{(a)}$, and J. Mateos Guilarte$^{(b)}$
\\ {\normalsize {\it $^{(a)}$ Departamento de Matematica
Aplicada and IUFFyM}, {\it Universidad de Salamanca, SPAIN}} \\ {\normalsize {\it $^{(b)}$ Departamento de Fisica
Fundamental and IUFFyM}, {\it Universidad de Salamanca, SPAIN}}}

\begin{document}

\maketitle

\begin{abstract}
In this paper we propose a generalization of the Gilkey-de Witt heat kernel expansion, designed to provide us with a precise estimation of the heat trace of non-negative Schr\"odinger type differential operators with non-trivial kernel over all the domain of its \lq\lq inverse temperature\rq\rq variable $\beta$. We apply this modified approach to compute effectively the one-loop kink mass shift for some models whose kink fluctuation operator spectrum is unknown and the only alternative to estimate this magnitude is the use of the heat kernel expansion techniques.
\end{abstract}


\section{Introduction}

In the mid sixties outstanding developments concerning the high-temperature expansion of the kernel of the generalized heat equation associated with a differential operator of the Laplace or Dirac type took place in different fields from Mathematics \cite{Gilkey1984, Roe1988} and Physics \cite{DeWitt1965}. In particular, Gilkey unveiled the meromorphic structure of the spectral zeta function, showing that the residua at the poles of this function are determined from the Seeley coefficients of the heat kernel expansion and describe certain topological invariants/characteristic classes. On the other hand De Witt used the heat kernel expansion to deal with quantum fields on curved backgrounds. Thereafter the heat kernel/zeta function methods have become an important tool in Quantum Field Theory \cite{Avramidi, Elizalde1994, Kirsten2002, Vassilevich2003}.

The computation of the one-loop kink mass correction is a topic enclosed in this framework. Technically this magnitude quantifies the contribution to the mass of the quantum fluctuations over the classical kink solution measured with respect to the background of vacuum quantum fluctuations, the kink Casimir energy. A mass renormalization must also be implemented in the previous scheme. All of this can be expressed in terms of the spectral zeta functions of the kink and vacuum fluctuation operators, which in turn can be written as an integral of its heat traces over the interval $\beta \in [0,\infty)$ by applying the Mellin transform. During the last decade we (and our colleagues) succeeded in applying the Gilkey-de Witt (GDW) heat kernel expansion in $(1+1)$-dimensional scalar field theoretical models to express the kink mass shift as a truncated series in the Seeley coefficients of the kink fluctuation operator heat trace, see \cite{Alonso2002,Alonso2011}. In some applications --heat kernel proofs of the index theorem, computation of anomalies in QFT-- the relevant information is encoded in the behavior of the heat trace for small values of the \lq\lq inverse temperature\rq\rq variable $\beta$. The kink mass shift, however, demands a precise estimation of the heat trace over all the domain of the variable, $\beta\in [0,\infty)$. We recall that the small $\beta$ regime captures the higher eigenvalues therefore determining the ultraviolet behavior while the large $\beta$ range is contrarily dominated by the lower eigenvalues and it is important in the infrared domain \cite{Alonso2012a}. The previous assertion is of no consequence if we dealt with strictly positive operators because in this case the GDW approach is well established; however it is crucial for the kink fluctuation operator, which always comprises a zero mode and consequently the GDW expansion does not reproduce the asymptotic behavior of the heat trace adequately. The usual manoeuvre to solve partially this problem is to truncate the Mellin transform in the range where there exists a good fitting between the GDW expansion and the heat trace. This obviously leads to a loss of precision in the final computations.

In this paper we introduce a different alternative; we modified the GDW route to accommodate the role of the zero modes in the theory. This is described in Section \S.2. In Section \S.3 we shall apply this modified GDW heat kernel expansion to obtain a precise computation of the one-loop kink mass shift for a (1+1) dimensional scalar field theoretical model, where other more direct methods (DHN formula, see \cite{Alonso2012}), are unapproachable because the kink fluctuation operator spectral information is unknown.

\section{The modified heat kernel asymptotic expansion}

\subsection{Heat kernel asymptotic expansion and zero modes}

In this section we shall address the generalization of the GDW heat kernel asymptotic expansion for non-negative second order differential operator of the form
\begin{equation}
K=-\frac{d^2}{dx^2}+v^2+V(x) \hspace{1cm}, \hspace{1cm} x\in \mathbb{R} \hspace{0.5cm},
\label{operatorK}
\end{equation}
where $V(x)$ is a real function whose asymptotic behavior complies with
\[
\lim_{x\rightarrow \pm \infty} V(x) =0\hspace{0.5cm}.
\]
In this case the $K$-spectrum
\[
{\rm Spec}(K)=\{\omega_0^2 = 0\} \cup \{\omega_n^2\}_{n=1,\dots,\ell} \cup \{k^2 + v^2\}_{k\in \mathbb{R}}
\]
embraces a zero mode $f_0(x)$, $\ell$ bound states $f_n(x)$ with non-negative eigenvalues and scattering states $f_k(x)$ emerging on the threshold value $v^2$. The computation of the heat integral kernel by means of its definition
\begin{equation}
K_{K}(x,y;\beta)= f_0^*(y) f_0(x)+\sum_{n=1}^\ell f_n^*(y)f_n(x) e^{-\beta \omega_n^2}+ \int \! dk \, f_k^{*} (y) \, f_k(x) \, e^{-\beta \omega^2(k)}
\label{integralkernel}
\end{equation}
is not possible for the most of the cases because we generally lack the spectral information of the $K$-operator. An alternative way to estimate $K_K(x,y;\beta)$ is to exploit the fact that (\ref{integralkernel}) is a solution of the parabolic (heat) equation
\begin{equation}
\left[ \frac{\partial}{\partial \beta}-\frac{\partial^2}{\partial
x^2}+v^2 + V(x) \right] K_{K}(x,y;\beta)=0 \hspace{0.4cm}, \label{heateq}
\end{equation}
with the contour conditions
\begin{equation}
\lim_{\beta\rightarrow +\infty} K_K(x,y;\beta)= f_0^*(y) f_0(x) \hspace{0.5cm},\hspace{0.5cm} \lim_{\beta\rightarrow 0} K_K(x,y;\beta)= \delta(x-y)\hspace{0.5cm},
\label{asympintkernel}
\end{equation}
dictated by the definition (\ref{integralkernel}). Therefore the zero mode rules the asymptotic behavior of $K_K(x,y;\beta)$. This constitutes a variation from the original GDW approach, where the second hand side of the first relation in (\ref{asympintkernel}) vanishes because of the positiveness assumption upon the $K$-spectrum assumed in that framework.

The potential wells of the $K$-operator (\ref{operatorK}) and the Helmholtz operator
\begin{equation}
K_0=-\frac{d^2}{dx^2}+v^2\hspace{1cm}, \hspace{1cm} x\in \mathbb{R} \hspace{0.5cm}, \label{operatorK00}
\end{equation}
have the same asymptotic behavior. The spectrum of this latter operator ${\rm Spec}(K_0)=\{k^2+v^2\}_{k\in \mathbb{R}}$ only comprises scattering states $f_k^0(x)=\frac{1}{\sqrt{2\pi}} e^{ikx}$. From this the exact $K_0$-heat kernel is
\begin{equation}
K_{K_0}(x,y;\beta)=\frac{1}{\sqrt{4\pi \beta\, }} \,
e^{-\beta v^2}\, e^{- \frac{(x-y)^2}{4 \, \beta\,}} \hspace{0.4cm},
\label{heatkernel0}
\end{equation}
whose contour conditions verify
\begin{equation}
\lim_{\beta\rightarrow +\infty} K_{K_0}(x,y;\beta)= 0 \hspace{0.5cm},\hspace{0.5cm} \lim_{\beta\rightarrow 0} K_{K_0}(x,y;\beta)= \delta(x-y)  \hspace{0.5cm}.
\label{asympintkernel0}
\end{equation}
The original GDW method analyzes the $K$-heat kernel $K_K(x,y,\beta)$ by means of its deviation from the exact form (\ref{heatkernel0}) using the standard factorization
\begin{equation}
K_{K}(x,y;\beta)=K_{K_0}(x,y;\beta) \, A(x,y;\beta) \quad  \mbox{with} \quad A(x,y;0)=1 \quad .
\label{factorization0}
\end{equation}
The relation (\ref{factorization0}) is, however, inconsistent in the present context because the two sides of (\ref{factorization0}) have different $\beta\rightarrow \infty$ asymptotic behaviors, see (\ref{asympintkernel}) and (\ref{asympintkernel0}). The reason of this underlies the presence of the zero mode in the $K$-spectrum. Notice, however, that (\ref{factorization0}) can be used to study the heat kernel $K_K(x,y;\beta)$ in the small $\beta$ range. This make possible, for instance, the heat kernel proofs of the index theorem, the computation of anomalies in QFT, etc. Nevertheless, other applications, such as the computation of the one-loop kink mass correction, demand the integration over the interval $\beta\in [0,\infty)$ of the heat kernel trace. Consequently the standard factorization (\ref{factorization0}) must be modified to reproduce the behavior of this function in the large $\beta$ range.

\subsection{The modified Gilkey-de Witt heat kernel expansion}

For non-negative differential operators of the type (\ref{operatorK}) the factorization (\ref{factorization0}) must be replaced by the form
\begin{equation}
K_{K}(x,y;\beta)=K_{K_0}(x,y;\beta) \, C(x,y;\beta) +g(\beta) e^{-\frac{(x-y)^2}{4\beta}} f_0^*(y) f_0(x)\, \, \, ,
\label{factorization}
\end{equation}
in order to adapt the GDW formalism to the existence of zero modes $f_0(x)$. The expression (\ref{factorization}) complies with the contour conditions (\ref{asympintkernel}) provided that
\begin{equation}
\lim_{\beta \rightarrow 0}C(x,y;0)=1 \hspace{1cm} , \hspace{1cm} \lim_{\beta \rightarrow \infty} g(\beta)=1 \hspace{0.4cm} , \hspace{0.4cm} \lim_{\beta \rightarrow 0}g(\beta)=0  \,\,\, . \label{heatic2}
\end{equation}
Plugging the ansatz (\ref{factorization}) into (\ref{heateq}) leads to the \lq\lq transfer" equation for $C(x,y;\beta)$:
\begin{eqnarray}
\label{heateq2.all}
&&0=\left( \frac{\partial}{\partial \beta}+\frac{x-y}{\beta}
\frac{\partial}{\partial x}-\frac{\partial^2}{\partial
x^2}+V(x) \right) C(x,y;\beta) +  \label{heateq2.a} \\ && + \sqrt{4\pi \beta} \, e^{\beta v^2}\, f_0^*(y) \left[ \frac{dg(\beta)}{d\beta} f_0(x)+ \frac{g(\beta)}{2\beta} f_0(x) +\frac{g(\beta)}{\beta} (x-y) \frac{df_0(x)}{dx} \right] \label{heateq2.b} \quad .
\end{eqnarray}
The terms specified in the line (\ref{heateq2.b}) concern to the role of the zero mode in this new scheme and complement the standard GDW \lq\lq transfer" equation (written in the line (\ref{heateq2.a})). The PDE (\ref{heateq2.all}) is traditionally solved by means of a power series expansion:
\begin{equation}
C(x,y;\beta) = \sum_{n=0}^\infty c_n(x,y) \, \beta^n  \quad \mbox{with} \quad c_0(x,y)=1 \quad ,
\label{expansion}
\end{equation}
where the constraint $c_0(x,y)=1$ is derived from the first condition in (\ref{heatic2}). Plugging the expression (\ref{expansion}) into (\ref{heateq2.all}) the relation
\begin{eqnarray}
& & \sum_{n=0}^\infty \left[ (n+1) c_{n+1}(x,y) - \frac{\partial^2 c_n(x,y)}{\partial x^2} +(x-y) \frac{c_{n+1}(x,y)}{\partial x} +V(x) c_n(x,y) \right] \beta^n + \nonumber \\
& &  \hspace{0.5cm} +\sqrt{4\pi \beta} e^{\beta v^2} f_0^*(y) \left[ \frac{dg(\beta)}{d\beta} f_0(x) + \frac{g(\beta)}{2\beta} f_0(x) + (x-y) \frac{g(\beta)}{\beta} \frac{df_0(x)}{dx} \right] = 0  \label{relation01}
\end{eqnarray}
holds. A significant simplification of (\ref{relation01}) is obtained by setting
\begin{equation}
\sqrt{\pi\beta} e^{\beta v^2} \frac{dg(\beta)}{d\beta}=v \hspace{0.5cm} \Rightarrow \hspace{0.5cm} g(\beta)={\rm Erf}\,(v\sqrt{\beta})  \,\,\, ,
\label{functiong}
\end{equation}
which complies with the asymptotic conditions (\ref{heatic2}) imposed on the function $g(\beta)$. Now the GDW procedure can be followed by implementing the asymptotic expansion of the error function
\[
{\rm Erf}\, z = \frac{2}{\sqrt{\pi}} e^{-z^2} \sum_{n=0}^\infty \frac{2^n}{(2n+1)!!}z^{2n+1}
\]
in the expression (\ref{relation01}). This leads to the recurrence relations for the coefficients $c_n(x,y)$:
\begin{eqnarray}\label{recursive1.all}
&& \hspace{0.8cm} 0=(n+1) \, c_{n+1}(x,y)-\frac{\partial^2 c_n(x,y)}{\partial
x^2} +(x-y) \frac{\partial c_{n+1}(x,y)}{\partial
x}+V(x) c_n(x,y) + \label{recursive1.a} \\ && \hspace{1cm}+ 2vf_0^*(y)f(x) \delta_{0n} + f_0^*(y)f(x) \frac{2^{n+1}v^{2n+1}}{(2n+1)!!} + (x-y)f_0^*(y) \frac{df_0(x)}{dx} \frac{2^{n+2}v^{2n+1}}{(2n+1)!!}\quad , \label{recursive1.b}
\end{eqnarray}
where again the terms displayed in (\ref{recursive1.b}) describe the new contributions to the standard expression (\ref{recursive1.a}). From the factorization (\ref{factorization}), the power series expansion (\ref{expansion}) and the choice (\ref{functiong}) of $g(\beta)$, the heat kernel can be written as the series expansion:
\begin{equation}
K_{K}(x,y;\beta)=K_{K_0}(x,y;\beta) \,\sum_{n=0}^\infty c_n(x,y) \, \beta^n +{\rm Erf}(\beta) e^{-\frac{(x-y)^2}{4\beta}} f_0^*(y) f_0(x)\quad ,
\label{facto2}
\end{equation}
where we recall that $f_0(x)$ is the zero mode of the $K$-operator and the coefficients $c_n(x,y)$ are computed by using the recurrence relation (\ref{recursive1.all}) starting from $c_0(x,y)=1$.

\subsection{The modified GDW heat trace expansion}

From the diagonal of the heat kernel
\begin{equation}
K_K(x,x,\beta)=\lim_{y\rightarrow x} K_K(x,y,\beta)= \frac{e^{-\beta v^2
}}{\sqrt{4 \pi \beta\,}} \sum_{n=0}^\infty c_n(x,x) \,
\beta^n + {\rm Erf}(\beta) f_0^*(x) f_0(x) \hspace{0.3cm},
\label{factorization7}
\end{equation}
we can derive the spectral $K$-heat trace $h_K(\beta)={\rm Tr}_{L^2}\, e^{-\beta K}$ as
\begin{equation}
h_K(\beta)=\int_\Omega \, dx \, K_K(x,x;\beta)  \hspace{0.3cm}. \label{heatkernel6}
\end{equation}
The Seeley densities $c_n(x,x)$ introduced in (\ref{factorization7}) are defined as
\begin{equation}
c_n(x,x)=\lim_{y\rightarrow x} c_n(x,y) \hspace{0.3cm}.
\end{equation}
From (\ref{recursive1.all}) a recursive relation can be constructed for these new coefficients. To do this it is necessary to deal with the following subtlety: the operations of taking the $y\rightarrow x$ limit and the derivatives with respect to $x$ in the formula (\ref{recursive1.all}) do not commute. To cope with this problem we introduce the new auxiliary coefficients:
\begin{equation}
{^{(k)}C}_n(x)=\lim_{y \rightarrow x} \frac{\partial^k
c_n(x,y)}{\partial x^k} \quad ,
\label{newcoef}
\end{equation}
whose first coefficients ${^{(k)}C}_0(x)$, $k\in \mathbb{N}$ are
\begin{equation}
{^{(k)} C}_0(x)=\lim_{y\rightarrow x} \frac{\partial^k
c_0}{\partial x^k}= \delta^{k0} \quad .
\label{ini0}
\end{equation}
Taking the $k$-th derivative of (\ref{recursive1.all}) with respect to $x$ and later passing to the $y\to x$ limit, the rest of the coefficients ${^{(k)} C}_n(x)$ verifies:
\begin{eqnarray}
{^{(k)} C}_n(x)& =&\frac{1}{n+k} \left[ \rule{0cm}{0.6cm} \right.
{^{(k+2)} C}_{n-1}(x) - \sum_{j=0}^k {k \choose j}
\frac{\partial^j V}{\partial x^j}\, \, {^{(k-j)}
C}_{n-1}(x) - \nonumber \\ &&  -2v f_0(x) \frac{df_0^k(x)}{dx^k} \delta_{0,n-1} - f_0(x) \frac{df_0^k(x)}{dx^k} \frac{2^n v^{2n-1}}{(2n-1)!!} (1+2k) \left. \rule{0cm}{0.6cm} \right] \quad ,
\label{capitalAcoefficients}
\end{eqnarray}
which must be used in a recursive way until the Seeley densities in (\ref{factorization7})
\begin{equation}
c_n(x,x)={^{(0)}C}_n(x) \label{SeeleyDensities}
\end{equation}
are determined. The first three Seeley densities $c_n(x,x)$ derived from (\ref{capitalAcoefficients}) for the $K$-operator are listed:
\begin{eqnarray}
c_0(x,x)={^{(0)} C}_0(x)&=&1\,\,, \nonumber \\
c_1(x,x)={^{(0)} C}_1(x)&=&-V(x)\hspace{2.7cm} -4v f_0^2(x)\,\,, \label{Seley1} \\
c_2(x,x)={^{(0)} C}_2(x)&=& \underbrace{-\frac{1}{6} \, \frac{\partial^2 {V}}{\partial
x^2}+\frac{1}{2} \, (V(x))^2}_{\rm standard \,\,terms} \hspace{0.5cm} \underbrace{+\frac{4}{3}v^3 f_0^2(x)+4v f_0^2(x) V(x)}_{\rm new \,\,terms} \,\,, \nonumber
\end{eqnarray}
where we distinguish between the terms coming from the standard GDW approach and the new contributions due to the presence of the zero mode. Finally the modified GDW heat trace expansion is obtained by plugging (\ref{factorization7}) into (\ref{heatkernel6}),
\[
h_{K}(\beta) =\frac{e^{-\beta v^2
}}{\sqrt{4 \pi \beta\,}} \sum_{n=0}^\infty c_n(K) \,
\beta^n + {\rm Erf}\,(v\sqrt{\beta}) \quad , \quad c_n(K)=\int_\Omega  dx \, c_n(x,x) \quad .
\]
where the Seeley coefficients $c_n(K)$ are given by the spatial integral of the Seeley densities $c_n(x,x)$. Therefore, the first three Seeley coefficients are
\begin{eqnarray}
c_0(K)&=&l\,\,, \nonumber\\
c_1(K)&=&- \left< V(x) \right> \hspace{3.1cm} -4 v\,\,, \label{Seley2} \\
c_2(K)&=& \underbrace{- \frac{1}{6} \left< V''(x) \right> + \frac{1}{2} \left< (V(x))^2 \right>}_{\rm standard \,\,terms}  \hspace{0.4cm} \underbrace{+\frac{4}{3} v^3 + 4 v \left< V(x) f_0^2(x) \right>}_{\rm new \,\,terms}  \,\, ,\nonumber
\end{eqnarray}
where again we point out the new terms introduced in this context. Here $l$ is the length of the interval $\Omega$, which eventually we make to tend to infinity. In any case, subtraction of the $K_0$-heat trace from the $K$-heat trace expansion amounts to dropping the $c_0(K)$ coefficient, and we find:
\begin{equation}
h_{K}(\beta)-h_{K_0}(\beta) = \frac{e^{-\beta v^2
}}{\sqrt{4 \pi}} \sum_{n=1}^\infty c_n(K) \,
\beta^{n-\frac{1}{2}} + {\rm Erf}\,(v\sqrt{\beta})  \,\,\, .
\label{heatfunction3}
\end{equation}

\section{A physical application: computation of the one-loop kink mass shift}

\subsection{Classical field theory models and kinks}

The notion of kink arises in the (1+1)-dimensional relativistic scalar field theory context \cite{Rajaraman1982, Drazin1996, Manton2004}. In this framework the dynamics is governed by the action
\[
\tilde{S}[\psi]=\int \!\! \int\, dy^0dy^1 \, \left(\frac{1}{2}\frac{\partial\psi}{\partial y_\mu}\cdot \frac{\partial\psi}{\partial y^\mu}- \tilde{U}[\psi(y^\mu)] \right) \quad .
\]
Here, $\psi(y^\mu): \mathbb{R}^{1,1} \rightarrow \mathbb{R}$ is a real scalar field; $y^0=\tau $ and $y^1=y$ are local coordinates in ${\mathbb R}^{1,1}$, equipped with a metric tensor $g_{\mu\nu}={\rm diag}(1,-1)$, $\mu,\nu=0,1$. We shall work in a system of units where the speed of light is set to one, $c=1$, but we shall keep the Planck constant $\hbar$ explicit because we shall search for one-loop corrections, proportional to $\hbar$, to the classical kink masses.
In this system, the physical dimensions of fields and parameters are: $[\hbar]=[\tilde{S}]=M L$, $[y_\mu]=L$, $[\psi]=M^\frac{1}{2}L^\frac{1}{2}$, $[\tilde{U}]=ML^{-1}$. In this type of models we can always identify two special parameters, $m_d$ and $\gamma_d$, to be determined in each case, carrying the physical dimensions: $[m_d]=L^{-1}$ and $[\gamma_d]=M^{-\frac{1}{2}}L^{-\frac{1}{2}}$, which allow us to introduce the non-dimensional coordinates, fields and potential: $x_\mu=m_d y_\mu$, $x_0=t$, $x_1=x$, $\phi= \gamma_d \psi$, $U(\phi)=\frac{\gamma_d^2}{m_d^2} \tilde{U}(\psi)$. The action and the \lq\lq static" part of the energy are also proportional to dimensionless action and energy functionals, namely:
\begin{eqnarray}
\tilde{S}[\psi]&=&\frac{1}{\gamma_d^2}S[\phi]= \frac{1}{\gamma_d^2}\int\!\!\!\int\, dx^0dx^1 \, \left[\frac{1}{2}\frac{\partial\phi}{\partial x_\mu}\cdot \frac{\partial\phi}{\partial x^\mu}- U[\phi(x^\mu)] \right] \, \, \, , \label{action}\\\tilde{E}[\psi]&=&\frac{m_d}{\gamma_d^2}E[\phi]= \frac{m_d}{\gamma_d^2}\int\!\! dx \, \left[\frac{1}{2}\left(\frac{d\phi}{d x}\right)^2 +U[\phi(x)] \right] \, \, \, , \label{energy}
\end{eqnarray}
where we shall assume that $U(\phi)$ is a non-negative twice-differentiable function of $\phi$: $U(\phi)\in C^2({\mathbb{R}})$ and $U(\phi)\geq 0$ for $\phi \in \mathbb{R}$. The configuration space ${\cal C}$ of the system is the set of field configurations ${\cal C}=\{\phi(t_0,x)\in {\rm Maps}(\mathbb{R}^{1},\mathbb{R})/ E[\phi]<+\infty\}$. The static and homogeneous solutions of the field equations
\begin{equation}
\left(\frac{\partial^2}{\partial t^2}-\frac{\partial^2}{\partial x^2}\right)\phi(t,x)=-\frac{\delta U}{\delta\phi}(t,x) \label{socfe}
\end{equation}
correspond to the vacua of the model, ${\cal M}=\{\phi^{(i)}\,\,/ \,\,U(\phi^{(i)})=0\}$. Topological solutions, kinks in this context, are defined as localized non-singular solutions of the field equation whose energy
density, as well as being localized, has space–time dependence of the form: $\varepsilon (t, x) = \varepsilon(x - vt)$, where $v$ is some velocity vector. The static kink solution $\phi_K(x)$ is a BPS solution connecting two vacua $\phi^{(i)}$ and $\phi^{(i+1)}$, which satisfies the first-order ODE
\begin{equation}
\frac{d\phi_K}{dx} =\pm \sqrt{2U(\phi_K)} \hspace{0.3cm}.
\label{edo5}
\end{equation}
Small perturbations around the kink and vacuum solutions are described by the spectral problem associated with the vacuum and kink fluctuation operators
\[
K_0=-\frac{d^2}{dx^2} + \frac{\partial^2 U}{\partial \phi^2}[\phi^{(i)}] \hspace{0.5cm},\hspace{0.5cm} K=-\frac{d^2}{dx^2} + \frac{\partial^2 U}{\partial \phi^2}[\phi_K(x)]
\]
which adopt respectively the form (\ref{operatorK00}) and (\ref{operatorK}) provided that we fix
\[
v^2=\frac{\partial^2 U}{\partial \phi^2}[\phi^{(i)}] \hspace{0.5cm},\hspace{0.5cm} V(x)=\frac{\partial^2 U}{\partial \phi^2}[\phi_K(x)]-v^2
\]

\subsection{One-loop kink mass shift and the zeta function regularization}

Dashen, Hasslacher and Neveu (DHN) solved the problem of computing the shift in the kink mass induced by kink fluctuations in the one-loop order of the sine-Gordon and $\lambda(\phi^4)_{1+1}$ models in \cite{Dashen1974}. The authors wrote the one-loop kink mass shift as the sum of two contributions $\bigtriangleup E(\phi_K)  = \bigtriangleup E_1(\phi_K) + \bigtriangleup E_2(\phi_K)$: the kink Casimir energy (mode-by-mode subtraction of the zero point vacuum energy) and the mass renormalization counter-term. In order to apply the DHN procedure effectively, it is necessary to know the eigenvalues of the bound states and the scattering wave phase shifts of the $K$ operator. For a generic scalar field theory model we lack this spectral information. Alternatively a zeta function regularization can be used in the previous scheme, see \cite{Alonso2002,Alonso2011,Bordag2002, Guilarte2009}. The vacuum energy induced by quantum fluctuations is first regularized by assigning it the value of the spectral zeta function of the $K_0$-operator (a meromorphic function) at a regular point in $s\in{\mathbb C}$:
\begin{equation}
\bigtriangleup E(\phi^{(i)})=\frac{\hbar \gamma_d^2}{2}\zeta_{K_0}(-{\textstyle\frac{1}{2}}) \, \, \rightarrow \, \, \bigtriangleup E(\phi^{(i)})[s]=\frac{\hbar \gamma_d^2}{2}\frac{\mu}{m_d}\left(\frac{\mu^2}{m_d^2}\right)^s\zeta_{K_0}(s) \quad ,
\end{equation}
where $\mu$ is a parameter of dimensions $L^{-1}$ introduced to keep the dimensions of the regularized energy right. We stress that $s=-\frac{1}{2}$ is a pole of this function. The same rule is applied to control the kink energy divergences by means of the spectral zeta function of $K$ and therefore the kink Casimir energy $\bigtriangleup E_1(\phi_K)[s]$ is regularized
in the form:
\begin{equation}
\hspace{0.5cm} \bigtriangleup E_1(\phi_K)[s]= \bigtriangleup E_0(\phi_K)[s]-\bigtriangleup E_0(\phi^{(i)})[s]=\frac{\hbar \gamma_d^2}{2}\left(\frac{\mu^2}{m_d^2}\right)^{s+\frac{1}{2}}\left(\zeta_{K}(s)-\zeta_{K_0}(s)\right) \,\,\, , \label{zrkc}
\end{equation}
On the other hand the energy due to the one-loop mass counter-term $\bigtriangleup E_2(\phi_K)[s]$ can be also regularized by the zeta function procedure:
\begin{equation}
\bigtriangleup E_2(\phi_K)[s]=  \frac{\hbar \gamma_d^2}{2}\langle V(x) \rangle \left(\frac{\mu^2}{m_d^2}\right)^{s+\frac{1}{2}}\lim_{l\to\infty}\frac{1}{l}\frac{\Gamma(s+1)}{\Gamma(s)}\zeta_{K_0}(s+1)
 \,\,\, .\label{zrkc3}
\end{equation}
Finally, we write the zeta function regularized DHN formula:
\begin{equation}
\bigtriangleup E(\phi_K)= \lim_{s\rightarrow -\frac{1}{2}} \bigtriangleup E_1(\phi_K)[s]+ \lim_{s\rightarrow -\frac{1}{2}} \bigtriangleup E_2(\phi_K)[s]  \,\,\, .
\label{zrkc4}
\end{equation}
The Mellin transform expresses the relation (\ref{zrkc}) in terms of the kink and vacuum fluctuation operator heat kernels:
\begin{equation}
\bigtriangleup E_1(\phi_K)[s]=\frac{\hbar \gamma_d^2}{2} \left(\frac{\mu^2}{m_d^2}\right)^{s+\frac{1}{2}}\frac{1}{\Gamma(s)}\left[\int_0^\infty \, d\beta \, \beta^{s-1}\left(h_{K}(\beta)-h_{K_0}(\beta)\right)\right] \,\,\, ,
\label{zrkc2}
\end{equation}
which allows us to apply the modified GDW approach introduced in the previous section. This leads to
\begin{eqnarray*}
\bigtriangleup E_1(\phi_K)[s]&=& \frac{\hbar \gamma_d^2}{2} \left( \frac{\mu^2}{m_d^2} \right)^{s+\frac{1}{2}} \Big[ -\frac{1}{\sqrt{4\pi}} \frac{\left<V(x)\right>}{v^{1+2s}} \frac{\Gamma[s+\frac{1}{2}]}{\Gamma[s]} - \frac{2}{\sqrt{\pi}\,v^{2s} } \frac{\Gamma[s+\frac{1}{2}]}{\Gamma[s]} + \\ &+& \frac{1}{\sqrt{4\pi}} \sum_{n=2}^\infty \frac{c_n(K)}{v^{2n+2s-1}} \frac{\Gamma[s+n-\frac{1}{2}]}{\Gamma[s]} - \frac{1}{\sqrt{\pi}v^{2s}} \frac{\Gamma[s+\frac{1}{2}]}{s\Gamma[s]} \, \Big] \quad ,
\end{eqnarray*}
where we have used the explicit expression (\ref{Seley2}) of $c_1(K)$ to write the contribution of the first term of the series. Because the contribution of the regularized one-loop mass counter-term $\bigtriangleup E_2(\phi_K)[s]$, see (\ref{zrkc3}), can be written as
\[
\bigtriangleup E_2(\phi_K)[s]=\frac{\hbar\gamma_d^2}{2}\left(\frac{\mu^2}{m_d^2}\right)^{s+\frac{1}{2}}
\frac{\langle V(x) \rangle}{\sqrt{4\pi}}\frac{\Gamma[s+\frac{1}{2}]}{\Gamma[s]}\frac{1}{v^{2s+1}}  \,\,\, ,
\]
we end from (\ref{zrkc4}) with the renormalized one-loop mass shift formula derived from the modified asymptotic series expansion of the $K$-heat function:
\begin{equation}
\frac{\bigtriangleup E(\phi_K)}{\hbar \gamma_d^2}=-\frac{v}{\pi}  -\frac{1}{8\pi} \sum_{n=2}^\infty c_n(K)(v^2)^{1-n}\Gamma[n-1]  \,\,\, .
\label{computation1}
\end{equation}
Either because of computational restrictions in the estimation of the Seeley coefficients or because of the asymptotic nature of the series, (\ref{computation1}) is truncated to a finite number $N_t$ of terms, which provides us with
\begin{equation}
\frac{\bigtriangleup E(\phi_K;N_t)}{\hbar \gamma_d^2}=-\frac{v}{\pi}  -\frac{1}{8\pi} \sum_{n=2}^{N_t} c_n(K)(v^2)^{1-n}\Gamma[n-1]
\label{numericalcorrection}
\end{equation}
as a good estimation of the quantum kink mass correction. The most important advantage of this procedure is that there is no need for detailed information about the spectrum of $K$ to calculate the mass shift.

\subsection{An example: the $\sinh^4 \phi$ model}

The potential term, which determines the dynamics in this model, is given by:
\[
U(\phi)=\frac{1}{4}(\sinh^2 \phi-1)^2 \quad .
\]
This function has two absolute minima located at $\phi^{(1)}=-{\rm arcsinh}\,1$ and $\phi^{(2)}={\rm arcsinh}\,1$, which play the role of vacua of the model. From the first order ODE (\ref{edo5}) we can identify the static kink solitary wave
\begin{equation}
\phi_K(x)={\rm arctanh} \frac{\tanh x}{\sqrt{2}}
\label{kink66}
\end{equation}
which connects the two vacuum points of the model. The fluctuations over this solution can be described by the spectral problems associated with the vacuum and kink Hessian operators, which are respectively:
\[
K_0=-\frac{d^2}{dx^2}+4 \hspace{1cm},\hspace{1cm} K=-\frac{d^2}{dx^2}+2+ \frac{16}{(1+{\rm sech}^2x)^2} - \frac{14}{1+{\rm sech}^2 x} \hspace{0.4cm}.
\]
The previous operators follow the form (\ref{operatorK}) and (\ref{operatorK00}) established in the first section, simply taking $v^2=4$ and $V(x)=-\frac{2\,{\rm sech}^2 x(9+{\rm sech}^2x)}{(1+{\rm sech}^2 x)^2}$. The one-loop mass shift for the kink can be computed by the formula (\ref{numericalcorrection}) where the zero mode $f_0(x)$, the normalized spatial derivative of the kink (\ref{kink66}), is $f_0(x)= \frac{4\sqrt{2}}{\sqrt{3\sqrt{2}\, {\rm arccosh}\, 3 -4} (3+\cosh (2x)}$ and the Seeley coefficients $c_n(K)$ are determined by means of the recurrence relations (\ref{capitalAcoefficients}). The Mathematica program KinkMassQuantumCorrection\_Modified.nb, which can be download at the web page http://campus.usal.es/$\sim$mpg/General/Mathematicatools, automatizes this calculation. The Seeley coefficients for the present case are listed in Table \ref{tab:results}.

\begin{table}[h]
  \caption{Seeley coefficients and partial kink mass shift estimations}
  \label{tab:results}
  \begin{center}\begin{tabular}{cccc}    \hline
      $n$ & $c_n(K)$ & $N_t$ & $\Delta E(\phi_K;N_t)/\hbar \gamma_d^2$    \\ \hline
      $1$ & $7.47870$ & $1$ & $-$ \\
      $2$ & $7.82708$ & $2$ & $-0.714477$ \\
      $3$ & $5.71901$ & $3$ & $-0.728699$ \\
      $4$ & $3.15228$ & $4$ & $-0.732619$ \\
      $5$ & $1.36104$ & $5$ & $-0.733888$ \\
      $6$ & $0.482077$ & $6$ & $-0.734338$ \\
      $7$ & $0.14436$ & $7$ & $-0.734506$ \\
      $8$ & $0.0375685$ & $8$ & $-0.734572$ \\
      \hline
  \end{tabular}\end{center}
\end{table}

In this Table we have also specified the values of the one-loop mass shift obtained from the modified asymptotic series (\ref{numericalcorrection}) for different values of the truncation order, $N_t$. These data have been depicted in Figure 1. For sake of comparison the same data extracted from the standard asymptotic expansion are also included, see \cite{Alonso2011}. Notice that the modified method exhibits better convergence properties than the standard procedure.

\begin{figure}[h]
\centerline{\includegraphics[height=3cm]{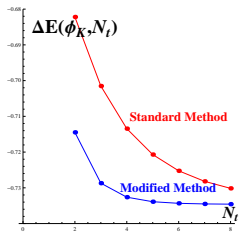} }
\caption{One-loop kink mass shifts estimated by means of the modified and standard asymptotic series for several truncation orders.}
\end{figure}

In conclusion, the one-loop mass shift for the kink (\ref{kink66}) in the $\sinh^4 \phi$ model is given by the value $-0.734572$, which has been obtained from the modified heat trace expansion truncated at the order $N_t=8$. The asymptotic behaviors of the partial sums associated with the modified and standard GDW heat trace expansions are illustrated in Figure 2. Notice that the modified approach reproduces the correct analytical asymptotic behavior as opposed to the standard procedure.

\begin{figure}[h]
\centerline{\includegraphics[height=2.7cm]{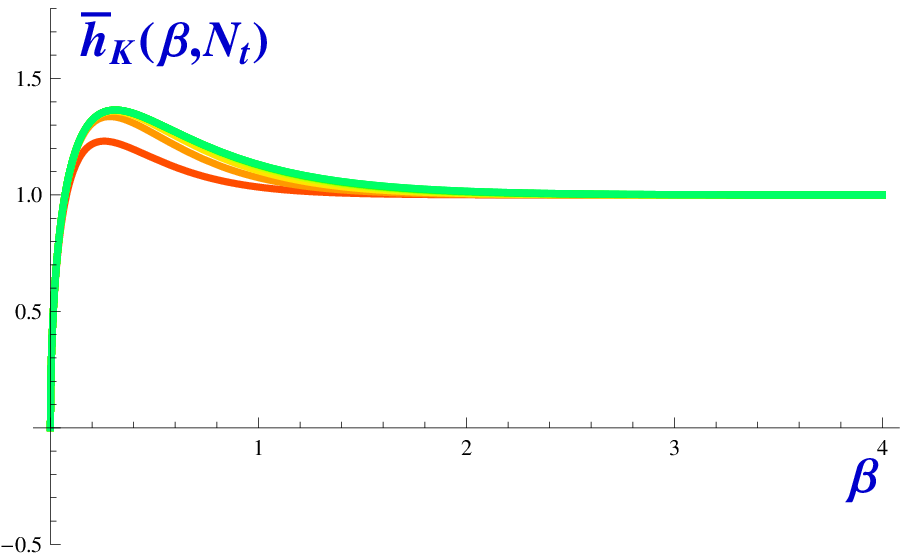} \hspace{1cm} \includegraphics[height=2.7cm]{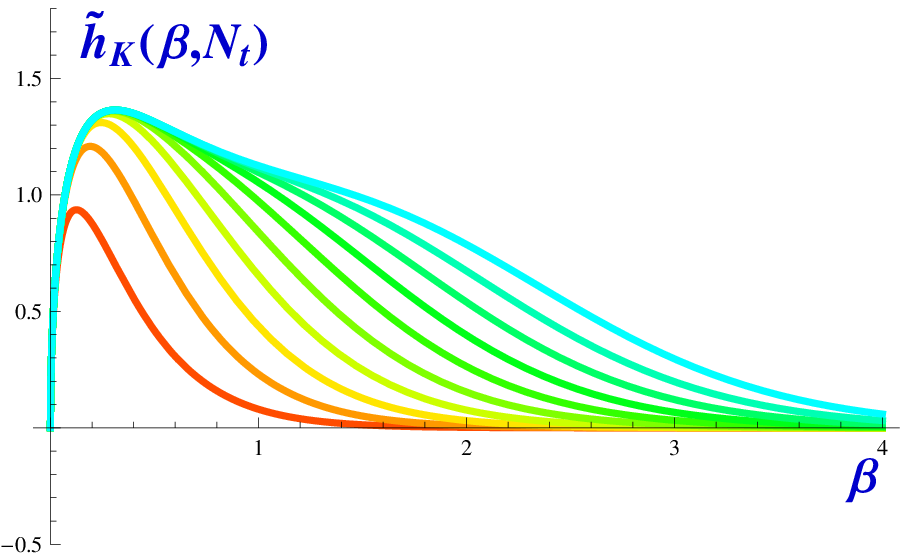} }
\caption{Graphics of the partial sums of the heat trace expansion from the modified (left) and standard (right) GDW approach.}
\end{figure}

\section*{Acknowledgments}

We warmly thank our collaborators in previous research into this topic, W. Garcia Fuertes, M. Gonzalez Leon and M. de la Torre Mayado. We also gratefully acknowledge conversations with D. Vassilevich and M. Bordag on this work.

\end{document}